\documentclass[preprintnumbers,prd,amsmath,amssymb,nofootinbib]{revtex4}
\usepackage{graphicx}
\usepackage{epsfig}
\usepackage{bm}
\usepackage{amsfonts}

\newcommand{\s}{\phi}
\newcommand{\la}{\lambda}

\newcommand{\be}{\begin{equation}}
\newcommand{\ee}{\end{equation}}

\begin{document}

\title{Phase-space analysis of interacting phantom cosmology}

\author{Xi-ming Chen} \email{chenxm@cqupt.edu.cn}
\affiliation{College of Mathematics and Physics,\\ Chongqing
University of Posts and Telecommunications, Chongqing 400065,
China}

\author{Yungui Gong} \email{gongyg@cqupt.edu.cn}
\affiliation{College of Mathematics and Physics,\\ Chongqing
University of Posts and Telecommunications, Chongqing 400065,
China}

\author{Emmanuel N.~Saridakis }
\email{msaridak@phys.uoa.gr} \affiliation{Department of Physics,\\
University of Athens,\\ GR-15771 Athens, Greece}

 \preprint{arXiv: 0812.1117}

\begin{abstract}
We perform a detailed phase-space analysis of various phantom
cosmological models, where the dark energy sector interacts with
the dark matter one. We examine whether there exist late-time
scaling attractors, corresponding to an accelerating universe and
possessing dark energy and dark matter densities of the same
order. We find that all the examined models, although accepting
stable late-time accelerated solutions, cannot alleviate the
coincidence problem, unless one imposes a form of fine-tuning in
the model parameters. It seems that interacting phantom cosmology
cannot fulfill the basic requirement that led to its construction.
 \end{abstract}

 \maketitle

\section{Introduction}

Recent cosmological observations support that the universe is
experiencing an accelerated expansion, and that the transition from
the deceleration phase to the accelerated phase happened in the
recent past \cite{observ}. In order to explain this remarkable
phenomena, one can modify the theory of gravity, such as $f(R)$
gravity \cite{invr,fr}, Dvali-Gabadadze-Porrati model \cite{dgp},
and string inspired models \cite{brane}. An alternative direction is
to introduce the concept of dark energy which provides the
acceleration mechanism. The simplest candidate of dark energy which
is consistent with current observations is the cosmological
constant. Due to the lack of a good explanation of the small value
of the cosmological constant, many dynamical dark energy models were
explored, such as a canonical scalar field (quintessence) model
\cite{quint}, a phantom model that has equation of state parameter
$w<-1$ \cite{phant,Boisseau:2000pr,ArmendarizPicon:2000ah}, or the
combination of quintessence and phantom in a unified model named
quintom \cite{quintom}. Moreover, many theoretical studies are
devoted to shed light on dark energy within some quantum
gravitational principles, such as the so-called ``holographic dark
energy'' proposal \cite{HOLO}.

The dynamical nature of dark energy introduces a new cosmological
problem, namely why are the densities of vacuum energy and dark
matter nearly equal today although they scale independently during
the expansion history. To solve the problem, one would require that
the matter density and dark energy density always approach their
current values independent of the initial conditions.  The
elaboration of this 'coincidence' problem led to the consideration
of generalized versions of the aforementioned models with the
inclusion of a coupling between dark energy and dark matter. Thus,
various forms of ``interacting'' dark energy models
\cite{interacting,Guo:2004xx} have been constructed in order to
fulfil the observational requirements.

One decisive test for dark energy models is the investigation of
their phase-space analysis. In particular, to examine whether they
possess attractor solutions corresponding to accelerating
universes and ratio $\Omega_{{\text{dark
energy}}}/\Omega_{{\text{dark matter}}}$ of the order 1. If these
conditions are fulfilled, then the universe will result to that
solution at late times, independently of the initial conditions,
and the basic observational requirements will be satisfied.
Although non-interacting quintessence
\cite{expon,Copeland:1997et}, non-interacting phantom
\cite{phannonin} and non-interacting quintom models
\cite{quinnonin} admit late-time accelerated attractors, they
possess $\Omega_{{\text{dark energy}}}=1$ and thus they are unable
to solve the coincidence problem.

When dark energy - dark matter interaction is switched on in
quintessence, then one can find accelerated attractors which
moreover give $\Omega_{{\text{dark energy}}}/\Omega_{{\text{dark
matter}}}\approx{\cal{O}}(1)$
\cite{Wetterich:1994bg,GarciaCompean:2007vh,Boehmer:2008av,Chen:2008pz},
but paying the price of introducing new problems such as
justifying a non-trivial, almost tuned, sequence of cosmological
epochs \cite{Amendola:2006qi}. In cases of interacting phantom
models \cite{Guo:2004vg}, the existing literature remains in some
special coupling forms with mainly numerical results, which
suggest that the coincidence problem might be alleviated
\cite{Guo:2004xx,Curbelo}.

In the present work we are interested in performing a detailed
phase-space analysis of the interacting phantom paradigm,
considering new and general forms of the interaction term $Q$. In
addition, we go beyond the simplified non-local forms of the
literature, considering also a local $Q$-form proportional to a
density
\cite{Boehmer:2008av,Cen:2000xv,Malik:2002jb,Ziaeepour:2003qs}.
Doing so we find that although late-time accelerated attractors do
exist in all the models, as it is expected for phantom cosmology,
almost all of them correspond to complete dark energy domination and
thus they are unable to solve the coincidence problem. Only for a
narrow area of the parameter space of one particular model, can the
corresponding solution alleviate the aforementioned problem.

The plan of the work is as follows: In section \ref{phantcosm} we
construct the interacting phantom cosmological scenario and we
present the formalism for its transformation into an autonomous
dynamical system, suitable for a phase-space stability analysis.
In section \ref{models} we perform the phase-space analysis for
four different models, using various interaction terms, and in
section \ref{cosmimpl} we discuss the corresponding cosmological
implications. Finally, section \ref{conclusions} is devoted to the
summary of the obtained results.

\section{Phantom Cosmology}
\label{phantcosm}

Let us construct the interacting phantom cosmological paradigm.
Throughout the work we consider a flat Robertson-Walker metric:
\begin{equation}\label{metric}
ds^{2}=dt^{2}-a^{2}(t)d\bf{x}^2,
\end{equation}
with $a$ the scale factor.

The evolution equations for the phantom and dark matter density
(considered as dust for simplicity) are:
\begin{eqnarray}\label{eom}
\dot{\rho}_m+3H\rho_m=-Q\\
\dot{\rho}_\s+3H(\rho_\s+p_\s)=Q,
\end{eqnarray}
with $Q$ a general interaction term and $H$ the Hubble parameter.
Therefore, $Q>0$ corresponds to energy transfer from dark matter to
dark energy, while $Q<0$ corresponds to dark energy transformation
to dark matter. For simplicity, we take the usual phantom energy
density and pressure:
\begin{eqnarray}
 \rho_{\s}&=& -\frac{1}{2}\dot{\s}^{2} + V(\s)\\
 p_{\s}&=& - \frac{1}{2}\dot{\s}^{2} - V(\s),
\end{eqnarray}
where $V(\s)$ is the phantom potential. Equivalently, the phantom
evolution equation can be written as:
\begin{equation}\label{sddot}
\ddot{\s}+3H\dot{\s}-\frac{\partial
V(\s)}{\partial\s}=-\frac{Q}{\dot{\s}}.
\end{equation}
Finally, the system of equation closes by considering the
Friedmann equations:
\begin{equation}\label{FR1}
H^{2}=\frac{\kappa^{2}}{3}(\rho_{\s}+\rho_{m}),
\end{equation}
\begin{equation}\label{FR2}
\dot{H}=-\frac{\kappa^2}{2}\Big(\rho_{\s}+p_{\s}+\rho_{m}\Big),
\end{equation}
where we have set $\kappa^2\equiv 8\pi G$. Although we could
straightforwardly include baryonic matter and radiation in the
model, for simplicity reasons we neglect them.

In phantom cosmological models, the dark energy is attributed to the
phantom field, and its equation of state is given by
\begin{equation}
w_{\s}=\frac{p_\s}{\rho_\s}.
\end{equation}
Alternatively one could construct the equivalent uncoupled model
described by:
\begin{eqnarray}
\dot{\rho}_m+3H(1+w_{m,eff})\rho_m=0\\
\dot{\rho}_\s+3H(1+w_{\s,eff})\rho_\s=0,
\end{eqnarray}
where
\begin{eqnarray}
w_{m,eff}&=&{Q \over 3H\rho_m}\\
w_{\s,eff}&=&w_\s -{Q \over 3H \rho_\s}.
\end{eqnarray}
However, it is more convenient to introduce the ``total''  energy
density $\rho_{tot}\equiv\rho_m+\rho_\s$, obtaining:
\begin{equation}
\label{rhot}
 \dot{\rho}_{tot}+3 H(1+w_{tot})\rho_{tot}=0,
\end{equation}
with
\begin{equation}
w_{tot}=\frac{p_\s}{\rho_\s+\rho_m}=w_\s\Omega_\s,
\end{equation}
where
$\Omega_\s\equiv\frac{\rho_\s}{\rho_{tot}}\equiv\Omega_{{\text{dark
energy}}}$. Obviously, since $\rho_{tot}=3H^2/\kappa^2$,
(\ref{rhot}) leads to a scale factor evolution of the form
$a(t)\propto t^{2/(3(1+w_{tot}))}$, in the constant $w_{tot}$
case. However, in the late-time stationary solutions that we are
studying in the present work, $w_{tot}$ has reached to a constant
value and thus the above behavior is valid. Therefore, we conclude
that in such stationary solutions the condition for acceleration
is just $w_{tot}<-1/3$.

In order to perform the phase-space and stability analysis of the
phantom model at hand, we have to transform the aforementioned
dynamical system into its autonomous form
\cite{expon,Copeland:1997et}. This will be achieved by introducing
the auxiliary variables:
\begin{eqnarray}
&&x=\frac{\kappa\dot{\s}}{\sqrt{6}H},\nonumber\\
&&y=\frac{\kappa\sqrt{V(\s)}}{\sqrt{3}H},
\label{auxilliary}
\end{eqnarray}
together with $M=\ln a$. Thus, it is easy to see that for every
quantity $F$ we acquire $\dot{F}=H\frac{dF}{dM}$.
 Using these
variables we obtain:
\begin{equation}
 \Omega_{\s}\equiv\frac{\kappa^{2}\rho_{\s}}{3H^{2}}=-x^2+y^2,
 \label{Omegas}
\end{equation}
\begin{equation}\label{wss}
w_{\s}=\frac{-x^2-y^2}{-x^2+y^2},
\end{equation}
and
\begin{equation}\label{wtot}
w_{tot}=-x^2-y^2.
\end{equation}
We mention that relations (\ref{wss}) and (\ref{wtot}) are always
valid, that is independently of the specific state of the system
(they are valid in the whole phase-space and not only at the
critical points).
Finally, note that in the case of complete dark energy
domination, i.e $\rho_m\rightarrow0$ and $\Omega_\s\rightarrow1$,
we acquire $w_{tot}\approx w_\s<-1$, as expected to happen in
phantom-dominated cosmology.

The next step is the introduction of a specific ansatz for the
interaction term $Q$. In this case the equations of motion
(\ref{eom}), (\ref{sddot}), (\ref{FR1}) and (\ref{FR2}) can be
transformed to an autonomous system containing the variables $x$
and $y$ and their derivatives with respect to $M=\ln a$. The
consideration of various $Q$-ansatzes is performed in the next
section.

A final assumption must be made in order to handle the potential
derivative that is present in (\ref{sddot}). The usual assumption in
the literature is to assume an exponential potential of the form
\begin{equation}
V=V_0\exp(-\kappa\lambda\s),
\end{equation}
since exponential potentials are known to be significant in
various cosmological models \cite{expon,Copeland:1997et}. Note
that equivalently, but more generally, we could consider
potentials satisfying $\lambda=-\frac{1}{\kappa
V(\s)}\frac{\partial V(\s)}{\partial\s}\approx const$ (for example
this relation is valid for arbitrary but nearly flat potentials
\cite{Scherrer:2007pu}).

Having transformed the cosmological system into its autonomous
form:
\begin{equation}\label{eomscol}
\textbf{X}'=\textbf{f(X)},
\end{equation}
where $\textbf{X}$ is the column vector constituted by the auxiliary
variables, \textbf{f(X)} the corresponding  column vector of the
autonomous equations, and prime denotes derivative with respect to
$M=\ln a$, we extract its critical points $\bf{X_c}$  satisfying
$\bf{X}'=0$. Then, in order to determine the stability properties of
these critical points, we expand (\ref{eomscol}) around  $\bf{X_c}$,
setting $\bf{X}=\bf{X_c}+\bf{U}$ with $\textbf{U}$ the perturbations
of the variables considered as a column vector. Thus, for each
critical point we expand the equations for the perturbations up to
the first order as:
\begin{eqnarray}
\label{perturbation} \textbf{U}'={\bf{\Xi}}\cdot \textbf{U},
\end{eqnarray}
where the matrix ${\bf {\Xi}}$ contains the coefficients of the
perturbation equations.
 Thus,
for each critical point, the eigenvalues of ${\bf {\Xi}}$
determine its type and stability.

\section{Phase-space analysis} \label{models}

In the previous section we constructed the interacting phantom
cosmological model, with an arbitrary interaction term $Q$, and we
presented the formalism for its transformation into an autonomous
dynamical system, suitable for a phase-space stability analysis.
In this section we introduce four specific forms for $Q$ and we
perform a complete phase-space analysis.

\subsection{Interacting model 1}

This specific interacting model is characterized by a coupling of
the form \cite{Chimento:2003iea,Boehmer:2008av,Chen:2008pz}:
\begin{equation}\label{mod1}
Q=\alpha H\rho_m.
\end{equation}
Thus, inserting the auxiliary variables (\ref{auxilliary}) into the
equations of motion (\ref{eom}), (\ref{sddot}), (\ref{FR1}) and
(\ref{FR2}), and using (\ref{mod1}) we result in the following
autonomous system:
\begin{eqnarray}\label{auto}
&&x'=-
3x+\frac{3}{2}x (1-x^2-y^2)-\sqrt{\frac{3}{2}}\la\, y^2-\frac{\alpha}{2x}(1+x^2-y^2)\nonumber\\
&&y'=\frac{3}{2}y (1-x^2-y^2)- \sqrt{\frac{3}{2}}\la\, x y
\label{autonomous1}.
\end{eqnarray}
The critical points $(x_c,y_c)$ of the autonomous system
(\ref{autonomous1}) are obtained by setting the left hand sides of
the equations to zero. The real and physically meaningful (i.e
corresponding to $y>0$ and $0\leq\Omega_\s\leq1$) of them are
presented in table \ref{stability1}. In the same table we present
the necessary conditions for their existence.

 The $2\times2$ matrix ${\bf
{\Xi}}$ of the linearized perturbation equations writes:
\[{\bf {\Xi}}= \left[ \begin{array}{cc}
3\left[-1+\frac{(1-x_c^2-y_c^2)}{2}-x_c^2-\frac{\alpha}{3}+\frac{\alpha}{6x_c^2}(1+x_c^2-y_c^2)\right] & y_c\left[-\sqrt{6}\lambda-3x_c+\frac{\alpha}{x_c}\right]  \\
y_c\left[-\sqrt{\frac{3}{2}}\lambda-3x_c\right] &
\left[\frac{3(1-x_c^2-y_c^2)}{2}-3y_c^2-\sqrt{\frac{3}{2}}\lambda
x_c\right]
\end{array} \right].\]
 Therefore,
for each critical point of table \ref{stability1}, we examine the
sign of the real part of the eigenvalues of ${\bf {\Xi}}$, which
determines the type and stability of this specific critical point.
In table \ref{stability1} we present the results of the stability
analysis. In addition, for each critical point we calculate the
values of $w_{tot}$ (given by relation (\ref{wtot})), and of
$\Omega_\s$ (given by  (\ref{Omegas})). Thus, knowing $w_{tot}$ we
can express the acceleration condition $w_{tot}<-1/3$ in terms of
the model parameters.

\begin{figure}[ht]
\begin{center}
\mbox{\epsfig{figure=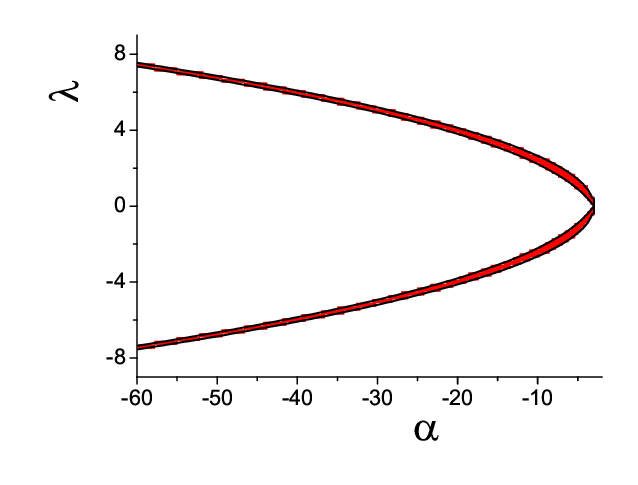,width=8cm,angle=0}}
\caption{{\it The range of the $2D$ parameter space
$(\alpha,\lambda)$ that corresponds to an existing and stable
critical point B of table \ref{stability1}, that is in the case of
interacting model 1. }} \label{Model1B}
\end{center}
\end{figure}

\begin{figure}[!]
\begin{center}
\includegraphics[width=8cm]{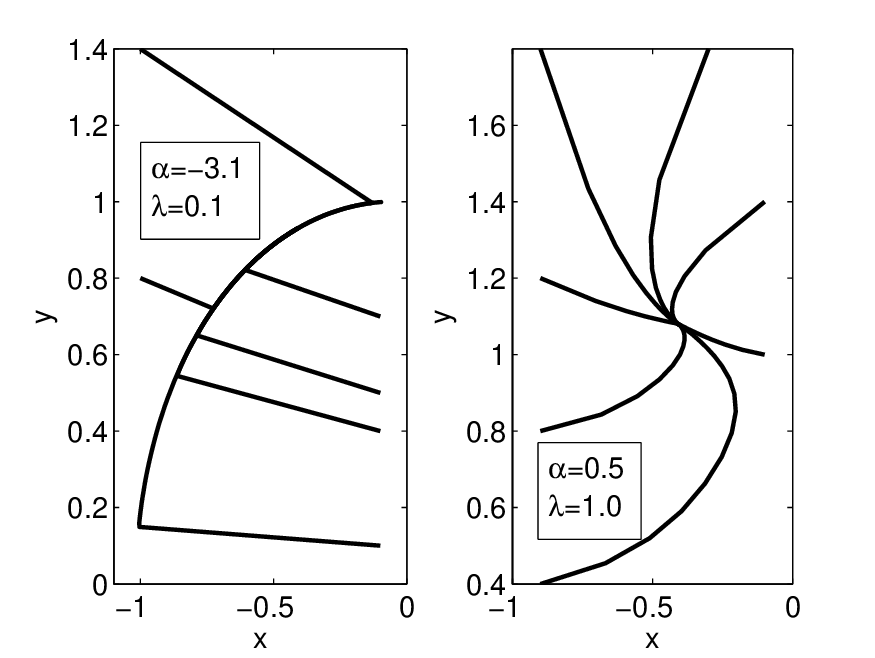}
\caption{Phase-space trajectories for interacting model 1. The
left panel corresponds to $\alpha=-3.1$ and $\lambda=0.1$, and
thus to the stable fixed point B of table \ref{stability1} with
($x_{c}$, $y_{c}$)=(-0.41, 0.93). The right panel corresponds to
$\alpha=0.5$ and $\lambda=1.0$ and thus to the stable fixed point
A of table \ref{stability1} with ($x_{c}$, $y_{c}$)=(-0.41,
1.08).} \label{cont1a}
\end{center}
\end{figure}

In order to present the results more transparently, in
fig.~\ref{Model1B} we depict the range of the $2D$ parameter space
$(\alpha,\lambda)$ that corresponds to an existing and stable
critical point B of table \ref{stability1}, i.e to negative real
parts of the corresponding eigenvalues. Furthermore, we
numerically evolve
 the autonomous system (\ref{auto}) for the parameters
$\alpha=0.5$, $\lambda=1.0$ and $\alpha=-3.1$, $\lambda=0.1$, and
the results are shown in figure \ref{cont1a}. Depending on which
region of the parameter-space do the chosen parameter-pair belong
(see table (\ref{stability1})), the system lies in the basin of
attraction of either the critical point A or B, and thus it is
attracted by the one or the other. This is apparent in figure
\ref{cont1a}, where we present two different parameter choices,
leading respectively to point B and A.
\begin{table*}[h]
\begin{center}
\begin{tabular}{|c|c|c|c|c|c|c|c|}
\hline
 Cr. P.& $x_c$ & $y_c$ & Existence & Stable for & $\Omega_\s$ &  $w_{tot}$ & Acceleration   \\
\hline \hline
 A&  $-\frac{\lambda}{\sqrt{6}}$ & $\sqrt{1+\frac{\lambda^2}{6}}$ & all $\alpha$,$\lambda$& $-\alpha<\lambda^2+3$ &1 &
  $-1-\frac{\lambda^2}{3}$ & all $\alpha$,$\lambda$ \\
\hline
 B&  $\frac{3+\alpha}{\sqrt{6}\lambda}$ & $\sqrt{-\frac{\alpha}{3}-\frac{(3+\alpha)^2}{6\lambda^2}}$ &
 $\lambda^2\geq\frac{(3+\alpha)^2}{-\alpha}$ for   $0\geq\alpha\geq-3$ & $\alpha<-3$&   $-\frac{(3+\alpha)^2}{3\lambda^2}-\frac{\alpha}{3}$
 &   $\frac{\alpha}{3}$ & $\alpha<-1$\\
 &  & &  $-(3+\alpha)\geq\lambda^2\geq\frac{(3+\alpha)^2}{-\alpha}$ for
 $\alpha\leq-3$ & $-(3+\alpha)>\lambda^2\geq\frac{(3+\alpha)^2}{-\alpha}$& & &\\
\hline
\end{tabular}
\end{center}
\caption[crit]{\label{stability1} The real and physically
meaningful critical points of  interacting model 1 and their
behavior.}
\end{table*}

\subsection{Interacting model 2}

In this model we consider an interaction term of the form
\cite{Chen:2008pz}:
\begin{equation}\label{mod2}
Q=\alpha_0 \kappa^{2n} H^{3-2n} \rho_m^n.
\end{equation}
This ansatz corresponds to a generalization of  interacting model
1. In the following, we take $n=2$ for simplicity. Inserting
(\ref{mod2}) into the cosmological equations (\ref{eom}),
(\ref{sddot}), (\ref{FR1}) and (\ref{FR2}), using the auxiliary
variables, we acquire the following autonomous system:
\begin{eqnarray}\label{autonomous2}
&&x'=-3x-\frac{\sqrt{6}}{2}\lambda
y^2+\frac{3}{2}x(1-x^2-y^2)-\frac{3}{2}\alpha_0\frac{(1+x^2-y^2)^2}{x}\nonumber\\
&&y'=-\frac{\sqrt{6}}{2}\lambda xy+\frac{3}{2}y(1-x^2-y^2) .
\end{eqnarray}
The real and physically meaningful critical points $(x_c,y_c)$ of
the autonomous system (\ref{autonomous2}) are:
 \be \label{mod2fxpts1}
\begin{split}
(x_{c1}=\pm \sqrt{\frac{-\alpha_0}{1+\alpha_0}},\ y_{c1}=0),
\quad (x_{c2}=-\frac{\lambda}{\sqrt{6}},\ y_{c2}=\sqrt{1+\frac{\lambda^2}{6}}),\\
(x_{c3},\ y_{c3}=\sqrt{1-x^2_{c3}-\sqrt{6}\lambda x_{c3}/3}),\quad
(x_{c4},\ y_{c4}=\sqrt{1-x^2_{c4}-\sqrt{6}\lambda x_{c4}/3}),
\end{split}
\ee
where
\be \label{xc4def}
x_{c3}=\frac{-(\alpha_0-1)\lambda +
\sqrt{\lambda^2(\alpha_0-1)^2-12\alpha_0}}{2\sqrt{6}\alpha_0},
 \ee
and
 \be \label{xc5def} x_{c4}=\frac{-(\alpha_0-1)\lambda -
\sqrt{\lambda^2(\alpha_0-1)^2-12\alpha_0}}{2\sqrt{6}\alpha_0}.
 \ee
These critical points are presented in table \ref{tab2}. The
matrix ${\bf {\Xi}}$ of the linearized perturbation equations is:
\[{\bf {\Xi}}= \left[ \begin{array}{cc}
3\left[-\frac{(1+3x_c^2+y_c^2)}{2}-2\alpha_0(1+x_c^2-y_c^2)+\frac{\alpha_0}{2}\frac{(1+x_c^2-y_c^2)^2}{x_c^2}\right] \quad & y_c\left[-\sqrt{6}\lambda-3x_c+6\alpha_0\frac{1+x_c^2-y_c^2}{x_c}\right]  \\
y_c\left[-\sqrt{\frac{3}{2}}\lambda-3x_c\right] &
\left[\frac{3(1-x_c^2-y_c^2)}{2}-3y_c^2-\sqrt{\frac{3}{2}}\lambda
x_c\right]
\end{array} \right].\]
Examining the eigenvalues of the matrix $\bf{\Xi}$ for each
critical point, we determine its stability behavior. Finally, in
table \ref{tab2} we also present
 $w_{tot}$,
$\Omega_\s$ and the acceleration condition. Note that the values
of $w_{tot}$ and $\Omega_\s$ for the critical points  C and  D can
be straightforwardly calculated using (\ref{mod2fxpts1}) and
(\ref{xc4def}), (\ref{xc5def}), but we do not present them since
these points are unstable and thus non-relevant for our
discussion.
\begin{table}[htp]
\begin{tabular}{|l|c|c|c|c|c|c|c|c|c|c|} \hline  Cr. P.&
$x_c$ & $y_c$ & Existence & Stable for & $\Omega_\phi$ &
$w_{tot}$ & Acceleration
\\
\hline \hline A & $\pm\sqrt{-\frac{\alpha_0}{1+\alpha_0}}$ & 0 &
$\alpha_0=0$ & Unstable  & $0$ & 0& No \\\hline
 B   &   $-\frac{\lambda}{\sqrt{6}}$ & $\sqrt{1+\frac{\lambda^2}{6}}$ & all
$\lambda$, $\alpha_0$ & all $\lambda$, $\alpha_0$ & 1 &
$-1-\frac{\lambda^2}{3}$ & all $\lambda$, $\alpha_0$
\\\hline
  C    &     $x_{c3}$ & $y_{c3}$ & {\tiny{$\lambda \le -\sqrt{\frac{3(1+\alpha_0)^2}{-\alpha_0}}$, $-1\le \alpha_0< 0$}} & Unstable & {\tiny{$-x_{c3}^2+y_{c3}^2$}} & {\tiny{$-x_{c3}^2-y_{c3}^2$}} &  {\tiny{ $-\sqrt{\frac{-4\alpha_0}{1+2\alpha_0}} < \lambda \le -\sqrt{\frac{3(1+\alpha_0)^2}{-\alpha_0}}$, $-1/2< \alpha_0< 0$}} \\
& & & {\tiny{ $\lambda \le \sqrt{\frac{3(1+\alpha_0)^2}{-\alpha_0}}$, $\alpha_0< -1$ }} & & & & $\lambda \le -\sqrt{\frac{3(1+\alpha_0)^2}{-\alpha_0}}$, $-1\le \alpha_0 \le -1/2$  \\
& & & & & & & $\lambda \le
\sqrt{\frac{3(1+\alpha_0)^2}{-\alpha_0}}$, $\alpha_0< -1$
\\\hline
 D   &   $x_{c4}$ & $y_{c4}$ & {\tiny{ $\lambda \ge \sqrt{\frac{3(1+\alpha_0)^2}{-\alpha_0}}$, $-1\le \alpha_0< 0$ }} & Unstable & {\tiny{  $-x_{c4}^2+y_{c4}^2$}} & {\tiny{ $-x_{c4}^2-y_{c4}^2$ }} &{\tiny{ $\sqrt{\frac{-4\alpha_0}{1+2\alpha_0}} > \lambda \ge \sqrt{\frac{3(1+\alpha_0)^2}{-\alpha_0}}$, $-1/2< \alpha_0< 0$}} \\
& & & {\tiny{ $\lambda \ge -\sqrt{\frac{3(1+\alpha_0)^2}{-\alpha_0}}$, $\alpha_0< -1$ }} & & & & $\lambda \ge \sqrt{\frac{3(1+\alpha_0)^2}{-\alpha_0}}$, $-1\le \alpha_0 \le -1/2$  \\
& & & & & & & $\lambda \ge
-\sqrt{\frac{3(1+\alpha_0)^2}{-\alpha_0}}$, $\alpha_0< -1$
\\\hline
\end{tabular}
 \caption{The real and physically
meaningful critical points of  interacting model 2 and their
behavior.} \label{tab2}
\end{table}

\begin{figure}[htp]
\centering
\includegraphics[width=8cm]{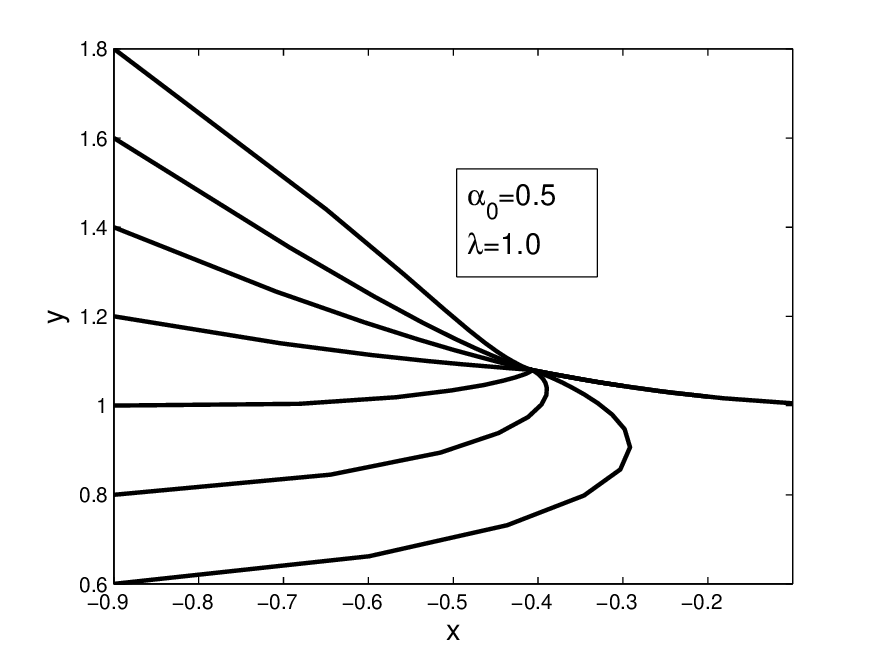}
\caption{Phase-space trajectories for  interacting model 2 using
$\alpha_0=0.5$ and $\lambda=1.0$. The stable fixed point is the
critical point $B$ in table \ref{tab2}, with ($x_{c2}$,
$y_{c2}$)=(-0.41, 1.08).} \label{cont2a}
\end{figure}

In order to make the phase-space analysis more transparent, we
elaborate the autonomous system (\ref{autonomous2}) numerically
for the parameter values $\alpha_0=0.5$ and $\lambda=1.0$. The
corresponding phase-space trajectories, arising imposing different
initial conditions, are shown in figure \ref{cont2a}. As expected,
the system is always attracted by the stable fixed point $B$ of
table \ref{tab2}.

\subsection{Interacting Model 3}

This model is characterized by an interaction term of the form
\begin{equation}\label{mod3}
Q=\beta \kappa^{2n} H^{1-2n}\rho_m^n \dot\phi^2,
\end{equation}
where for simplicity we consider $n=1$. The
 autonomous system reads:
\begin{eqnarray}\label{autonomous3}
&&x'=-3x-\frac{\sqrt{6}}{2}\lambda
y^2+\frac{3}{2}x(1-x^2-y^2)-3\beta x(1+x^2-y^2)\nonumber\\
&& y'=-\frac{\sqrt{6}}{2}\lambda xy+\frac{3}{2}y(1-x^2-y^2).
\end{eqnarray}
The real and physically meaningful critical points $(x_c,y_c)$ of
the autonomous system (\ref{autonomous3}) are:
 \be
\label{mod3fxpts1}
\begin{split}
(x_{c1}=0,\ y_{c1}=0),
\quad (x_{c2}=-\frac{\lambda}{\sqrt{6}},\ y_{c2}=\sqrt{1+\frac{\lambda^2}{6}}),\\
(x_{c3},\ y_{c3}=\sqrt{1-x^2_{c3}-\sqrt{6}\lambda x_{c3}/3}),\quad
(x_{c4},\ y_{c4}=\sqrt{1-x^2_{c4}-\sqrt{6}\lambda x_{c4}/3}),
\end{split}
\ee where \be \label{mod3xc3}
x_{c3}=\frac{\lambda+\sqrt{\lambda^2-12\beta}}{2\sqrt{6}\beta},
\ee and \be \label{mod3xc4}
x_{c4}=\frac{\lambda-\sqrt{\lambda^2-12\beta}}{2\sqrt{6}\beta}.
\ee The matrix ${\bf {\Xi}}$ of the linearized perturbation
equations is:
\[{\bf {\Xi}}= \left[ \begin{array}{cc}
3\left[-\frac{(1+3x_c^2+y_c^2)}{2}-\beta (1+3x_c^2-y_c^2)\right] \quad & y_c\left[-\sqrt{6}\lambda-3x_c+6\beta x_c\right]  \\
y_c\left[-\sqrt{\frac{3}{2}}\lambda-3x_c\right] &
\left[\frac{3(1-x_c^2-y_c^2)}{2}-3y_c^2-\sqrt{\frac{3}{2}}\lambda
x_c\right]
\end{array} \right].\]
 The critical points are presented in table \ref{tab3}, together
 with their stability behavior,
 $w_{tot}$,
$\Omega_\s$ and the acceleration condition.
\begin{table}[htp]
\begin{tabular}{|l|c|c|c|c|c|c|c|c|} \hline Cr. P.&
$x_c$ & $y_c$ & Existence & Stable for & $\Omega_\phi$ &
$w_{tot}$& Acceleration
\\
\hline\hline A & 0 & 0 & all $\beta$, $\lambda$ & Unstable & 0 & 0
& No
\\
\hline B  &
 $-\frac{\lambda}{\sqrt{6}}$ &
$\sqrt{1+\frac{\lambda^2}{6}}$ &all $\beta$, $\lambda$ & {\rm all}\ $\lambda$, $\beta\ge -1$ & 1 & $-1-\frac{\lambda^2}{3}$ & all $\beta$, $\lambda$ \\
 &  &  &  &   $\lambda^2<-\frac{3}{1+\beta}$, $\beta <-1$     & &  &  \\
 \hline
 C  & $x_{c3}$
& $y_{c3}$ & $\lambda \le \sqrt{\frac{-3}{1+\beta}}$, $\beta<
-1$ & Unstable & $-x_{c3}^2+y_{c3}^2$  &$-x_{c3}^2-y_{c3}^2$ &
$-\sqrt{-4\beta}<\lambda\le
\sqrt{\frac{-3}{1+\beta}}$, $\beta< -1$ \\
\hline
D   &
 $x_{c4}$ & $y_{c4}$ & $-\sqrt{\frac{-3}{1+\beta}}\le
\lambda$, $\beta< -1$ & Unstable & $-x_{c4}^2+y_{c4}^2$ &
$-x_{c4}^2-y_{c4}^2$  &
$-\sqrt{\frac{-3}{1+\beta}}\le\lambda<\sqrt{-4\beta}$, $\beta<
-1$ \\\hline
\end{tabular}
 \caption{ The real and physically
meaningful critical points of interacting model 3 and their
behavior.} \label{tab3}
\end{table}

\begin{figure}[ht]
\centering
\includegraphics[width=8cm]{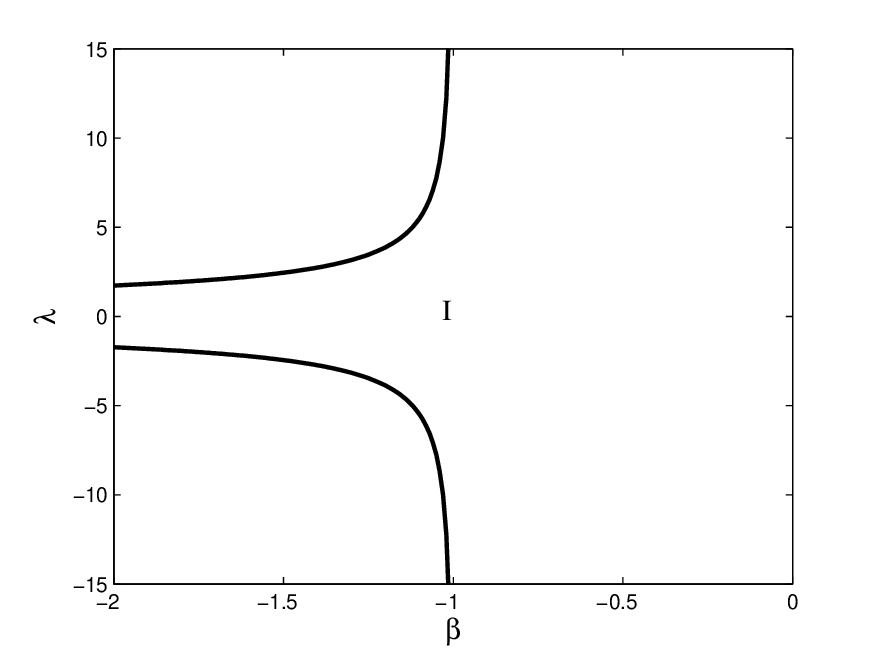}
\caption{The range of the $2D$ parameter space $(\beta,\lambda)$
that corresponds to an existing and stable critical point B of
table  \ref{tab3}, that is in the case of interacting model 3. }
\label{pht3par}
\end{figure}

\begin{figure}[!]
\centering
\includegraphics[width=8cm]{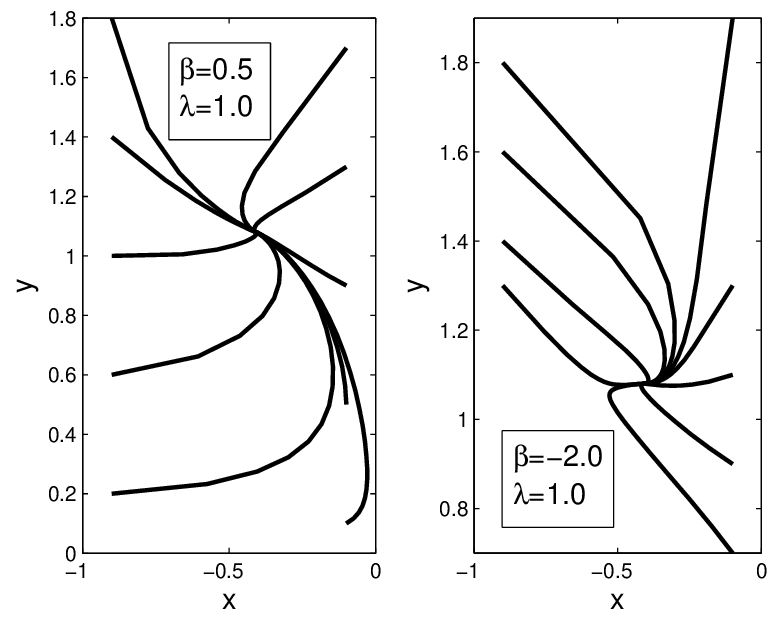}
\caption{Phase-space trajectories for  interacting model 3 with
$\beta=0.5$, $\lambda=1.0$  (left panel) and $\beta=-2.0$,
$\lambda=1.0$ (right panel). In both cases the stable fixed point
is the critical point B of table \ref{tab3}, with ($x_{c2}$,
$y_{c2}$)=(-0.41, 1.08). } \label{cont3a}
\end{figure}

In order to present the results more transparently, in
fig.~\ref{pht3par} we depict the range of the $2D$ parameter space
$(\beta,\lambda)$ that corresponds to the existence of the stable
critical point B of table \ref{tab3}. Furthermore, we elaborate
the autonomous system (\ref{autonomous3}) numerically, for the
parameter values $\lambda=1.0$,  $\beta=0.5$ and $\lambda=1.0$,
$\beta=-2.0$ . The
 phase-space trajectories, corresponding to different
initial conditions, are shown in figure \ref{cont3a}. In these
cases, the system is always attracted by the stable fixed point B
of table \ref{tab3}, as expected.

\subsection{Interacting Model 4}

In the previous subsections we considered interaction terms
depending explicitly on the Hubble parameter $H$, that is
interaction terms determined by the properties of the whole
universe. In the present  model we consider the local interaction
term \cite{Boehmer:2008av}
\begin{equation}\label{mod4}
Q=\Gamma\rho_m,
\end{equation}
where we assume that the coefficient $\Gamma$ is a constant.
 Note that this ansatz, with $\Gamma>0$, has been used in
 different frameworks to describe the decay of dark matter into
radiation \cite{Cen:2000xv}, the decay of a curvaton field into
radiation \cite{Malik:2002jb} and the decay of superheavy dark
matter particles into a quintessence scalar field
\cite{Ziaeepour:2003qs}.

The complexity of the interaction term (\ref{mod4}), comparing to
the previous ones, is revealed by the fact that the dynamical
evolution cannot be elaborated using only the two auxiliary
variables (\ref{auxilliary}). Indeed, $H$ cannot be eliminated from
the autonomous equations. In order to achieve that we introduce an
additional auxiliary variable as
\begin{equation}
v=\frac{H_0}{H},
 \label{zzz}
\end{equation}
where $H_0$ is a constant. Thus, using (\ref{mod4}), the
cosmological equations  (\ref{eom}), (\ref{sddot}), (\ref{FR1})
and (\ref{FR2}) are transformed to the following
 autonomous system:
\begin{eqnarray}
&&x'=-
3x+\frac{3}{2}x (1-x^2-y^2)-\sqrt{\frac{3}{2}}\la\, y^2-\frac{\gamma v}{2x}(1+x^2-y^2)\nonumber\\
&&y'=\frac{3}{2}y (1-x^2-y^2)- \sqrt{\frac{3}{2}}\la\, x y\nonumber\\
&&v'=\frac{3}{2}v(1-x^2-y^2) \label{autonomous4},
\end{eqnarray}
where we have introduced the dimensionless coupling constant
\begin{equation}
\gamma=\frac{\Gamma}{H_0}
 \label{gamma}.
\end{equation}
The $3\times 3$ matrix ${\bf {\Xi}}$ of the linearized
perturbation equations is:
\[{\bf {\Xi}}= \left[ \begin{array}{ccc}
3\left[-\frac{(1+3x_c^2+y_c^2)}{2}+\frac{\gamma v_c}{6x_c^2}(1+x_c^2-y_c^2)-\frac{\gamma v_c}{3}\right] \quad & y_c\left[-\sqrt{6}\lambda-3x_c+\frac{\gamma v_c}{x_c}\right] \quad
& -\frac{\gamma}{2x_c}(1+x_c^2-y_c^2)  \\
y_c\left[-\sqrt{\frac{3}{2}}\lambda-3x_c\right] &
\left[\frac{3(1-x_c^2-y_c^2)}{2}-3y_c^2-\sqrt{\frac{3}{2}}\lambda
x_c\right] & 0\\
-3x_c v_c & -3y_c v_c & \frac{3}{2}(1-x_c^2-y_c^2)
\end{array} \right].\]
In this model there is only one real and physically meaningful
critical point presented in table \ref{stability4}, together
 with its stability behavior,
 $w_{tot}$,
$\Omega_\s$ and the acceleration condition.
\begin{table*}[h]
\begin{center}
\begin{tabular}{|c|c|c|c|c|c|c|c|c|}
\hline
 Cr. P.& $x_c$ & $y_c$ &  $v_c$ & Existence & Stable for & $\Omega_\s$ &  $w_{tot}$ & Acceleration   \\
\hline \hline
 A&  $-\frac{\lambda}{\sqrt{6}}$ & $\sqrt{1+\frac{\lambda^2}{6}}$ &0 & all $\gamma$,$\lambda$ & all $\gamma$,$\lambda$ &1 &
  $-1-\frac{\lambda^2}{3}$ & all $\gamma$,$\lambda$ \\
 \hline
\end{tabular}
\end{center}
\caption[crit]{\label{stability4} The real and physically
meaningful critical point of interacting model 4 and its
behavior.}
\end{table*}
\begin{figure}[ht]
\begin{center}
\includegraphics[width=6.5cm]{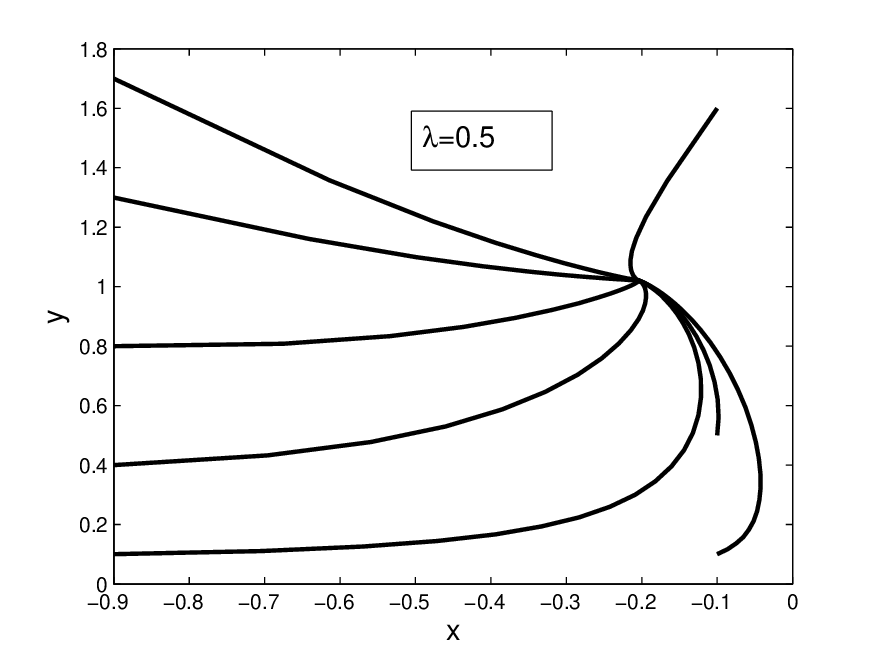}
\caption{ Phase-space trajectories for  interacting model 4 with
$\lambda=0.5$. The stable fixed points is the critical point A  of
table \ref{stability4}, with ($x_{c}$, $y_{c}$)=(-0.20, 1.02).}
\label{cont4a}
\end{center}
\end{figure}

To reveal this behavior more transparently, we solve numerically
the autonomous system (\ref{autonomous4}) for $\lambda=0.5$. In
order to plot the two dimensional phase trajectories for the
variables $x$ and $y$, we project the system onto the $v=0$ plane
and the results are shown in figure \ref{cont4a}. Finally, we
mention that for the corresponding model of a canonical field,
examined in \cite{Boehmer:2008av},  using the variable
$z=H_0/(H_0+H)$ the authors found  ``critical points'' for $z=1$,
which corresponds to $v=\infty$ in our case. However, as we can
see, this point is not a critical point of the system.

\section{Cosmological implications}
\label{cosmimpl}

Since we have performed a complete phase-space analysis of various
interacting phantom models, we can now discuss the corresponding
cosmological behavior. A general remark  is that this behavior is
radically different from the corresponding interacting
quintessence models with the same couplings
\cite{Chen:2008pz,Boehmer:2008av}. Furthermore, a common feature
of all the examined models is that we obtain $w_{\s}<-1$ in the
whole phase-space, and $w_{tot}<-1$ for all the existing and
stable critical points. Thus, independently of the specific
interacting model and of the chosen initial conditions, we will be
always moving on the same side (below) of the phantom divide. This
was expected for phantom cosmology. However, we remind that
interacting quintessence can conditionally describe the
$-1$-crossing, in contrast to  expectations
\cite{Wetterich:1994bg,GarciaCompean:2007vh,Boehmer:2008av,Chen:2008pz}.
Finally,  we mention that the fact that $w_{tot}$ is not only less
than $-1/3$, as required by the acceleration condition, but it is
always less than $-1$ in all stable solutions, leads
 to $\dot{H}>0$ at all times. Thus, these solutions correspond to a
super-accelerating universe \cite{Das:2005yj}, that is with a
permanently increasing $H$. If $H\rightarrow\infty$ at
$t\rightarrow\infty$ then we acquire an eternally expanding
universe, while if $H\rightarrow\infty$ at $t\rightarrow
t_{BR}<\infty$ then the universe results to a Big Rip. These
behaviors are also common in phantom cosmology
\cite{phant,Briscese:2006xu}.

\subsection{Interacting Model 1}

In this model, both critical points A and B are stable points, and
thus late-time attractors (see table \ref{stability1}). Point A,
which is stable for a large region of the $2D$ parameter space
($\alpha,\lambda$), corresponds to an accelerating universe. In
addition, since $\Omega_\s=1$ it corresponds to a complete dark
energy domination. Note that both these features hold
independently of the $\alpha$-value if it is positive, which was
expected since positive $\alpha$ means positive $Q$, that is
energy transfer from dark matter to dark energy. So it leads to a
final, complete dark energy domination. In conclusion, we see that
the stable fixed point A does correspond to an accelerating
universe, but cannot solve the coincidence problem since it leads
to $\Omega_{{\text{dark energy}}}=1$ instead of
$\Omega_{{\text{dark energy}}}/\Omega_{{\text{dark
matter}}}\approx{\cal{O}}(1)$.

On the other hand, the critical point B, for $\alpha<-3$ and for
$\lambda$-values that make it stable (note that in this case the
existence and stability conditions coincide) corresponds to an
accelerating universe with $0<\Omega_\s<1$, i.e with
$\Omega_{{\text{dark energy}}}/\Omega_{{\text{dark
matter}}}\approx{\cal{O}}(1)$. Therefore, for this region of the
($\alpha,\lambda$) parameter space, the stable fixed point B can
satisfy the observational requirements, and at late times the
universe is attracted by an accelerating solution with vacuum and
matter densities of the same order. For example, if we take
$\alpha=-3.1$ and $\lambda=0.1$ (see left panel of fig.
\ref{cont1a}), then we obtain  $\Omega_\phi=0.7$ and
$w_{tot}=-1.03$, which is in agreement with current observations
within $95\%$ confidence level \cite{observ}. However, as can be
seen from figure \ref{Model1B},  the stability region of point B,
which produces the aforementioned interesting cosmological
behavior, is very narrow. Therefore, the late-time attractor A has
significantly larger probability to be the universe's late-time
solution. In other words, the phantom interacting model at hand
can indeed satisfy the cosmological requirements, but for a very
narrow area of the parameter space.

\subsection{Interacting Model 2}

In this case we observe that the critical points A, C and D are
unstable, and therefore they cannot be late-time cosmological
solutions. The only relevant critical point is B, which is a stable
fixed points for the whole ($\alpha_0,\lambda$) parameter space. As
can be seen from table \ref{tab2}, it corresponds to an accelerating
universe with $\Omega_\s=1$, i.e., to a complete dark-energy
domination. Thus, it cannot act as a candidate for solving the
coincidence problem.

\subsection{Interacting Model 3}

In this model we see that the critical points A, C and D are
unstable and thus non-relevant. Critical point B is the only
cosmologically interesting one, since it corresponds to a late-time
attractor for the region of the ($\beta,\lambda$) parameter space
shown in figure \ref{pht3par}. However, although for that region it
always gives rise to acceleration, it possesses $\Omega_\s=1$.
Therefore it cannot solve the coincidence problem.

\subsection{Interacting Model 4}

In this case, the critical point A is the only real and physically
meaningful critical point, and it is stable for the whole $2D$
parameter space ($\gamma,\lambda$). It corresponds to an
accelerating universe, completely dominated by dark energy
($\Omega_\s=1$). Thus this phantom interacting model cannot solve
the coincidence problem.

\section{Conclusions}
\label{conclusions}

In this work we performed a detailed phase-space analysis of various
phantom cosmological models, where the dark energy sector interacts
with the dark matter one. Our basic goal was to examine whether
there exist late-time scaling attractors, corresponding to
accelerated universe and possessing $\Omega_{{\text{dark
energy}}}/\Omega_{{\text{dark matter}}}\approx{\cal{O}}(1)$, thus
satisfying the basic observational requirements. We investigated
four different interaction models, including one with a local, i.e.
$H$-independent, form of the interaction term (interacting model 4).
We extracted the critical points, determined their stabilities, and
calculated the basic cosmological observables, namely the total
equation-of-state parameter $w_{tot}$ and $\Omega_{{\text{dark
energy}}}$ (attributed to the phantom field). The key point of
solving the coincidence problem is that the universe is in the
attractor now and it has normal radiation dominated and matter
dominated eras before it reached the attractors. Note that as long
as the interaction term is not too strong, the standard cosmology
can be always recovered. Once it is in the attractor, the results do
not depend on the initial conditions. Thus, one can switch on the
interaction and consider as initial conditions the end of the known
epochs of standard Big Bang cosmology, in order to avoid disastrous
interference.

In all the examined models we found that stable late-time
solutions do exist, corresponding moreover  to an accelerating
universe. This feature was expected since phantom cosmology has
been constructed in order to always satisfy this condition.
Indeed, for all the studied models, we did not find any
non-accelerating stable solution. However, in almost all the cases
the late-time solutions correspond to a complete dark energy
domination and thus are unable to solve the coincidence problem.
The only case in which this is possible is in interacting model 1,
if we select the parameter values from a very narrow region of the
$2D$ parameter space (see figure \ref{Model1B}).

In conclusion, we see that the examined models of interacting
phantom cosmology can produce acceleration (which is ``embedded''
in phantom cosmology in general) but cannot solve the coincidence
problem, unless one imposes a form of fine-tuning in the model
parameters. This result has been extracted by the
negative-kinetic-energy  realization of phantom, which does not
cover the whole class of phantom models, but since it is a
qualitative statement it should intuitively be robust for general
interacting phantom scenarios, too. An alternative direction would
be to consider more complicated interaction terms, suitably
constructed in order to solve the coincidence problem. But the
interaction term was introduced in phantom cosmology in order to
solve the coincidence problem in a simple and general way,
avoiding the assumptions and fine-tunings of conventional
cosmology. Although promising and with many advantages,
interacting phantom cosmology needs further investigation.


\begin{thebibliography}{99}



\bibitem{observ}
A.G. Riess {\it et al.} [Supernova Search Team Collaboration],
Astron. J. {\bf 116}, 1009 (1998);
A. G. Riess {\it{et al.}}
[Supernova Search Team Collaboration], Astrophys. J. {\bf 607}, 665
(2004);
 S.
Perlmutter {\it{et al.}} [Supernova Cosmology Project
Collaboration], Astrophys. J. {\bf 517}, 565 (1999);
 D. N. Spergel {\it{et al.}}, Astrophys.
J. Suppl. {\bf 148}, 175 (2003); S. W. Allen, {\it{et al.}}, Mon.
Not. Roy. Astron. Soc. {\bf 353}, 457 (2004).


\bibitem{invr} S.M. Carroll, V. Duvvuri, M. Trodden, M.S. Turner, Phys. Rev. D {\bf 70}, 043528 (2004);
T. Chiba, Phys. Lett. B {\bf 575},  1 (2003); S. Nojiri, S.D.
Odintsov, Phys. Rev. D {\bf 68}, 123512 (2003); C.G. Shao, R.G.
Cai, B. Wang, R.K. Su, Phys. Lett. B {\bf 633}, 164 (2006).

\bibitem{fr} S. Capozziello, Int. J. Mod. Phys. D {\bf11}, 483 (2002);
%\cite{Appleby:2007vb}
  S.~A.~Appleby and R.~A.~Battye,
  %``Do consistent $F(R)$ models mimic General Relativity plus $\Lambda$?,''
  Phys.\ Lett.\  B {\bf 654}, 7 (2007);
  %%CITATION = PHLTA,B654,7;%%
S. Nojiri, S.D. Odintsov, Int. J. Geom. Meth. Mod. Phys. {\bf4},
115 (2007);
%\cite{Starobinsky:2007hu}
  A.~A.~Starobinsky,
  %``Disappearing cosmological constant in f(R) gravity,''
  JETP Lett.\  {\bf 86}, 157 (2007);
  %%CITATION = JTPLA,86,157;%%
 W. Hu, I. Sawicki, Phys. Rev. D {\bf76}, 064004
(2007); S. Nojiri, S.D. Odintsov, [arXiv:0807.0685[hep-th]].

\bibitem{dgp} G.R. Dvali, G. Gabadadze, M. Porrati, Phys. Lett. B {\bf485}, 208 (2000);
C. Deffayet, G.R. Dvali, G. Gabadadze, Phys. Rev. D {\bf65},
044023 (2002); Y.G. Gong, C.K. Duan, Class. Quantum Grav. {\bf21},
3655 (2004);
%\cite{Diakonos:2007au}
  F.~K.~Diakonos and E.~N.~Saridakis,
  %%CITATION = ARXIV:0708.3143;%%
 JCAP {\bf 0902}, 030 (2009);
Y.G. Gong, C.K. Duan, Mon. Not. Roy. Astron. Soc. {\bf352}, 847
(2004); Y.G. Gong, [arXiv:0808.1316[astro-ph]].

\bibitem{brane} P. Bin\'{e}truy, C. Deffayet, D. Langlois, Nucl. Phys. B {\bf565}, 269 (2000);
R.G. Cai, Y.G. Gong, B. Wang, JCAP {\bf0603}, 006 (2006); Y.G.
Gong, A. Wang, Class. Quantum Grav. {\bf23}, 3419 (2006); Y.G.
Gong, A. Wang, Q. Wu, Phys. Lett. B {\bf663}, 147 (2008);
%\cite{Setare:2008hm}
  M.~R.~Setare and E.~N.~Saridakis,
  %``Correspondence between Holographic and Gauss-Bonnet dark energy models,''
  Phys.\ Lett.\  B {\bf 670}, 1 (2008);
  %%CITATION = PHLTA,B670,1;%%;
  %%CITATION = ARXIV:0810.3296;%%
  %\cite{Setare:2008mb}
  M.~R.~Setare and E.~N.~Saridakis,
 JCAP {\bf 0903}, 002 (2009).
  %%CITATION = ARXIV:0811.4253;%%


\bibitem{quint}
B.~Ratra and P.~J.~E.~Peebles, Phys.\ Rev.\ D {\bf 37}, 3406
(1988); C.~Wetterich, Nucl.\ Phys.\ B {\bf 302}, 668 (1988);
A.~R.~Liddle and R.~J.~Scherrer, Phys.\ Rev.\ D {\bf 59}, 023509
(1998); I.~Zlatev, L.~M.~Wang and P.~J.~Steinhardt, Phys.\ Rev.\
Lett.\ {\bf 82}, 896 (1999); Z.~K.~Guo, N.~Ohta and Y.~Z.~Zhang,
Mod.\ Phys.\ Lett.\  A {\bf 22}, 883 (2007).

\bibitem{phant} R. R. Caldwell, Phys.
Lett. B {\bf{545}}, 23 (2002); R.~R.~Caldwell, M.~Kamionkowski and
N.~N.~Weinberg, Phys. Rev. Lett. {\bf 91}, 071301 (2003); S.
Nojiri and S. D. Odintsov, Phys. Lett. B {\bf 562}, 147 (2003); V.
K. Onemli and R. P. Woodard, Phys.\ Rev.\ D {\bf 70}, 107301
(2004) [arXiv:gr-qc/0406098]; M. R. Setare, Eur. Phys. J. C {\bf
50}, 991 (2007);
%\cite{Saridakis:2008fy}
  E.~N.~Saridakis,
  %``Theoretical Limits on the Equation-of-State Parameter of Phantom
  %Cosmology,''
  [arXiv:0811.1333 [hep-th]].
  %%CITATION = ARXIV:0811.1333;%%

\bibitem{Boisseau:2000pr}
  B.~Boisseau, G.~Esposito-Farese, D.~Polarski and A.~A.~Starobinsky,
  %``Reconstruction of a scalar-tensor theory of gravity in an accelerating
  %universe,''
  Phys.\ Rev.\ Lett.\  {\bf 85}, 2236 (2000);
  %%CITATION = PRLTA,85,2236;%%
%\cite{Nojiri:2005vv}
  S.~Nojiri, S.~D.~Odintsov and M.~Sasaki,
  %``Gauss-Bonnet dark energy,''
  Phys.\ Rev.\  D {\bf 71}, 123509 (2005);
  %%CITATION = PHRVA,D71,123509;%%
  %\cite{Li:2005fm}
  M.~z.~Li, B.~Feng and X.~m.~Zhang,
  %``A single scalar field model of dark energy with equation of state  crossing
  %-1,''
  JCAP {\bf 0512}, 002 (2005);
  %%CITATION = JCAPA,0512,002;%%
%\cite{Nojiri:2005sr}
  S.~Nojiri and S.~D.~Odintsov,
  %``Inhomogeneous equation of state of the universe: Phantom era, future
  %singularity and crossing the phantom barrier,''
  Phys.\ Rev.\  D {\bf 72}, 023003 (2005);
  %%CITATION = PHRVA,D72,023003;%%
  %\cite{Sur:2008tc}
  S.~Sur and S.~Das,
  %``Multiple kinetic k-essence, phantom barrier crossing and stability,''
  JCAP {\bf 0901}, 007 (2009);
  %%CITATION = JCAPA,0901,007;%%
%\cite{Bamba:2008hq}
  K.~Bamba, C.~Q.~Geng, S.~Nojiri and S.~D.~Odintsov,
  %``Crossing of the phantom divide in modified gravity,''
  arXiv:0810.4296 [hep-th].
  %%CITATION = ARXIV:0810.4296;%%

%\cite{ArmendarizPicon:2000ah}
\bibitem{ArmendarizPicon:2000ah}
  C.~Armendariz-Picon, V.~F.~Mukhanov and P.~J.~Steinhardt,
  %``Essentials of k-essence,''
  Phys.\ Rev.\  D {\bf 63}, 103510 (2001)
  [arXiv:astro-ph/0006373].
  %%CITATION = PHRVA,D63,103510;%%


\bibitem{quintom}
B.~Feng, X.~L.~Wang and X.~M.~Zhang, Phys.\ Lett.\  B {\bf 607},
35 (2005);
  %%CITATION = PHLTA,B607,35;%%
Z. K. Guo, {\it{et al.}}, Phys. Lett. B {\bf 608}, 177 (2005);
M.-Z Li, B. Feng, X.-M Zhang, JCAP,  {\bf0512}, 002 (2005); B.
Feng, M. Li, Y.-S. Piao and X. Zhang, Phys. Lett. B {\bf 634}, 101
(2006); M. R. Setare, Phys. Lett. B {\bf 641}, 130 (2006); W. Zhao
and Y. Zhang, Phys. Rev. D {\bf73}, 123509 (2006);
  %%CITATION = JCAPA,0809,026;%%
%\cite{Setare:2008pc}
  M.~R.~Setare and E.~N.~Saridakis,
  %``Non-minimally coupled canonical, phantom and quintom models of holographic
  %dark energy,''
  Phys.\ Lett.\  B {\bf 671}, 331 (2009).
  %%CITATION = PHLTA,B671,331;%%



\bibitem{HOLO}
  A.~G.~Cohen, D.~B.~Kaplan and A.~E.~Nelson,
  Phys.\ Rev.\ Lett.\  {\bf 82}, 4971 (1999);
  P.~Horava and D.~Minic,
  Phys.\ Rev.\ Lett.\  {\bf 85}, 1610 (2000);
  S.~D.~H.~Hsu,
  Phys.\ Lett.\ B {\bf 594}, 13 (2004);
  M.~Li,
  Phys.\ Lett.\ B {\bf 603}, 1 (2004);
  Y.G. Gong, Phys. Rev. D  {\bf70}, 064029 (2004);
   Y.G. Gong and J. Liu, JCAP  {\bf0809},
010 (2008);
  D.~Pavon and W.~Zimdahl,
  Phys.\ Lett.\ B {\bf 628}, 206 (2005);
  H.~Li, Z.~K.~Guo and Y.~Z.~Zhang,
  Int.\ J.\ Mod.\ Phys.\ D {\bf 15}, 869 (2006);
M. R. Setare, J. Zhang and X. Zhang, JCAP {\bf 0703}, 007 (2007);
%\cite{Saridakis:2007cy}
  E.~N.~Saridakis,
  Phys.\ Lett.\  B {\bf 660}, 138 (2008);
  %%CITATION = PHLTA,B660,138;%%
 %\cite{Saridakis:2007ns}
  E.~N.~Saridakis,
  JCAP {\bf 0804}, 020 (2008);
  %%CITATION = JCAPA,0804,020;%%
 %\cite{Saridakis:2007wx}
  E.~N.~Saridakis,
  Phys.\ Lett.\  B {\bf 661}, 335 (2008).
  %%CITATION = PHLTA,B661,335;%%

\bibitem{interacting}
 A. P. Billyard and A. A. Coley, Phys. Rev D {\bf 61}, 083503 (2000); J. P. Mimoso, A.
Nunes and D.Pavon, Phys. Rev. D {\bf 73}, 023502 (2006); R. Lazkoz
and G. Leon, Phys. Lett. B {\bf 638}, 303 (2006);
 T.~Gonzalez, G.~Leon and I.~Quiros,
  %``Dynamics of Quintessence Models of Dark Energy with Exponential Coupling to
  %the Dark Matter,''
  Class.\ Quant.\ Grav.\  {\bf 23}, 3165 (2006);
  %[arXiv:astro-ph/0702227].
  %%CITATION = CQGRD,23,3165;%%
  G.~R.~Farrar and P.~J.~E.~Peebles,
  Astrophys.\ J.\  {\bf 604}, 1 (2004);
%\cite{Wang:2005jx}
  B.~Wang, Y.~G.~Gong and E.~Abdalla,
  %``Transition of the dark energy equation of state in an interacting
  %holographic dark energy model,''
  Phys.\ Lett.\  B {\bf 624}, 141 (2005);
  %%CITATION = PHLTA,B624,141;%%
%\cite{Setare:2006wh}
  M.~R.~Setare,
  %``Interacting holographic dark energy model in non-flat universe,''
  Phys.\ Lett.\  B {\bf 642}, 1 (2006).
  %%CITATION = PHLTA,B642,1;%%

%\cite{Guo:2004xx}
\bibitem{Guo:2004xx}
  Z.~K.~Guo, R.~G.~Cai and Y.~Z.~Zhang,
  JCAP {\bf 0505}, 002 (2005);
A. Nunes, J.P. Mimoso and T.C. Charters, Phys. Rev. D {\bf 63},
083506 (2001); D.F. Mota and C. van de Bruck, Astron. Astrophys.
{\bf 421}, 71 (2004); M. Manera and D.F. Mota, Mon. Not. Roy.
Astron. Soc. {\bf 371}, 1373 (2006); N.J. Nunes and D.F. Mota, Mon.
Not. Roy. Astron. Soc. {\bf 368}, 751 (2006); J.D. Barrow and T.
Clifton, Phys. Rev. D {\bf 73}, 103520 (2006); T. Clifton and J.D.
Barrow, Phys. Rev. D {\bf 73}, 104022 (2006); T. Clifton and J.D.
Barrow, Phys. Rev. D {\bf 75}, 043515 (2007); M. Jamil and M.A.
Rashid, arXiv: 0802.1144; M. Jamil, arXiv: 0810.2896.

\bibitem{expon} P.G. Ferreira, M. Joyce, Phys. Rev. Lett.  {\bf79}, 4740 (1997);
 E.J. Copeland, M. Sami, S. Tsujikawa, Int. J. Mod. Phys. D
 {\bf15}, 1753 (2006); Y.G. Gong, A. Wang, Y.Z. Zhang, Phys. Lett. B
 {\bf636}, 286
(2006).


%\cite{Copeland:1997et}
\bibitem{Copeland:1997et}
  E.~J.~Copeland, A.~R.~Liddle and D.~Wands,
  %``Exponential potentials and cosmological scaling solutions,''
  Phys.\ Rev.\  D {\bf 57}, 4686 (1998).
  %%CITATION = PHRVA,D57,4686;%%



\bibitem{phannonin}
  P.~Singh, M.~Sami and N.~Dadhich,
  Phys.\ Rev.\  D {\bf 68}, 023522 (2003);
  J.~G.~Hao and X.~Z.~Li,
  Phys.\ Rev.\  D {\bf 70}, 043529 (2004).


\bibitem{quinnonin}
  H.~Wei and S.~N.~Zhang,
  Phys.\ Rev.\  D {\bf 76}, 063005 (2007);
  %\cite{Setare:2008si}
  M.~R.~Setare and E.~N.~Saridakis,
  JCAP {\bf 0809}, 026 (2008);
  %%CITATION = JCAPA,0809,026;%%
%\cite{Setare:2008pz}
  M.~R.~Setare and E.~N.~Saridakis,
  Phys.\ Lett.\  B {\bf 668}, 177 (2008);
  %%CITATION = PHLTA,B668,177;%%
%\cite{Setare:2008dw}
  M.~R.~Setare and E.~N.~Saridakis,
  %``Quintom Cosmology with General Potentials,''
  [arXiv:0807.3807 [hep-th]].
  %%CITATION = ARXIV:0807.3807;%%




\bibitem{Wetterich:1994bg}
  C.~Wetterich,
  Astron.\ Astrophys.\  {\bf 301}, 321 (1995);
  L.~Amendola,
  Phys.\ Rev.\  D {\bf 60}, 043501 (1999).


%\cite{GarciaCompean:2007vh}
\bibitem{GarciaCompean:2007vh}
  H.~Garcia-Compean, G.~Garcia-Jimenez, O.~Obregon and C.~Ramirez,
  %``Crossing the phantom divide in an interacting generalized Chaplygin gas,''
  JCAP {\bf 0807}, 016 (2008).
  %%CITATION = JCAPA,0807,016;%%



%\cite{Chen:2008pz}
\bibitem{Chen:2008pz}
  X.~M.~Chen and Y.G. Gong,
  %``Fixed points in interacting dark energy models,''
  [arXiv:0811.1698[gr-qc]].
  %%CITATION = ARXIV:0811.1698;%%


%\cite{Boehmer:2008av}
\bibitem{Boehmer:2008av}
  C.~G.~B\"{o}hmer, G.~Caldera-Cabral, R.~Lazkoz and R.~Maartens,
  %``Dynamics of dark energy with a coupling to dark matter,''
  Phys.\ Rev.\  D {\bf 78}, 023505 (2008).
  %%CITATION = PHRVA,D78,023505;%%


\bibitem{Amendola:2006qi}
  L.~Amendola, M.~Quartin, S.~Tsujikawa and I.~Waga,
  Phys.\ Rev.\  D {\bf 74}, 023525 (2006).

%\cite{Guo:2004vg}
\bibitem{Guo:2004vg}
  Z.~K.~Guo and Y.~Z.~Zhang,
  %``Interacting phantom energy,''
  Phys.\ Rev.\  D {\bf 71}, 023501 (2005).
  %%CITATION = PHRVA,D71,023501;%%
%\cite{Gonzalez:2007ht}
  T.~Gonzalez and I.~Quiros,
  %``Exact models with non-minimal interaction between dark matter and (either
  %phantom or quintessence) dark energy,''
  Class.\ Quant.\ Grav.\  {\bf 25}, 175019 (2008).
  %%CITATION = CQGRD,25,175019;%%





\bibitem{Curbelo}
R.~Curbelo, T.~Gonzalez, G. Leon and I.~Quiros,
  %``Interacting phantom energy and avoidance of the big rip singularity,''
  Class.\ Quant.\ Grav.\  {\bf 23}, 1585 (2006).
        %%CITATION = CQGRD,23,1585;%%

%\cite{Cen:2000xv}
\bibitem{Cen:2000xv}
  R.~Cen,
  %``Decaying Cold Dark Matter Model and Small-Scale Power,''
  Astrophys.\ J.\  {\bf 546}, L77 (2001)
  [arXiv:astro-ph/0005206];
%\cite{Oguri:2003nn}
%\bibitem{Oguri:2003nn}
  M.~Oguri, K.~Takahashi, H.~Ohno and K.~Kotake,
  %``Decaying Cold Dark Matter and the Evolution of the Cluster Abundance,''
  Astrophys.\ J.\  {\bf 597}, 645 (2003).

%\cite{Malik:2002jb}
\bibitem{Malik:2002jb}
  K.~A.~Malik, D.~Wands and C.~Ungarelli,
  %``Large-scale curvature and entropy perturbations for multiple interacting
  %fluids,''
  Phys.\ Rev.\  D {\bf 67}, 063516 (2003).

%\cite{Ziaeepour:2003qs}
\bibitem{Ziaeepour:2003qs}
  H.~Ziaeepour,
  %``Quintessence from the decay of a superheavy dark matter,''
  Phys.\ Rev.\  D {\bf 69}, 063512 (2004).


%\cite{Boisseau:2000pr}


\bibitem{Scherrer:2007pu}
  R.~J.~Scherrer and A.~A.~Sen,
  Phys.\ Rev.\  D {\bf 77}, 083515 (2008);
  R.~J.~Scherrer and A.~A.~Sen,
  Phys.\ Rev.\  D {\bf 78}, 067303 (2008);
%\cite{Setare:2008sf}
  M.~R.~Setare and E.~N.~Saridakis,
  %``Quintom dark energy models with nearly flat potentials,''
  Phys.\ Rev.\  D {\bf 79}, 043005 (2009).
  %%CITATION = ARXIV:0810.4775;%%

\bibitem{Chimento:2003iea}
  L.~P.~Chimento, A.~S.~Jakubi, D.~Pavon and W.~Zimdahl,
  Phys.\ Rev.\  D {\bf 67}, 083513 (2003).

%\cite{Das:2005yj}
\bibitem{Das:2005yj}
  S.~Das, P.~S.~Corasaniti and J.~Khoury,
  Phys.\ Rev.\  D {\bf 73}, 083509 (2006);
  M.~Kaplinghat and A.~Rajaraman,
  Phys.\ Rev.\  D {\bf 75}, 103504 (2007).
  %%CITATION = PHRVA,D75,103504;%%

%\cite{Briscese:2006xu}
\bibitem{Briscese:2006xu}
  F.~Briscese, E.~Elizalde, S.~Nojiri and S.~D.~Odintsov,
  %``Phantom scalar dark energy as modified gravity: Understanding the origin of
  %the big rip singularity,''
  Phys.\ Lett.\  B {\bf 646}, 105 (2007).
  %%CITATION = PHLTA,B646,105;%%


\end{thebibliography}
\end{document}